\documentclass[11pt]{article}

\usepackage{a4wide}
\usepackage{latexsym}
\usepackage{graphicx}
\usepackage{color}
\usepackage{hyperref}

\title{A Model for Collaboration Networks Giving Rise to a Power Law Distribution with an Exponential Cutoff}

\author{Trevor Fenner, Mark Levene, and George Loizou \\
School of Computer Science and Information Systems \\
Birkbeck College, University of London \\
London WC1E 7HX, U.K. \\ \{trevor,mark,george\}@dcs.bbk.ac.uk}

\date{}

\begin{document}

\maketitle

\newtheorem{theorem}{Theorem}[section]
\newtheorem{corollary}[theorem]{Corollary}
\newtheorem{lemma}[theorem]{Lemma}
\newtheorem{proposition}[theorem]{Proposition}
\newtheorem{definition}{Definition}[section]
\newtheorem{algorithm}{Algorithm}
\newtheorem{example}{Example}[section]

\begin{abstract}

Recently several authors have proposed stochastic evolutionary models for the growth
of complex networks that give rise to power-law distributions. These models are based
on the notion of preferential attachment leading to the ``rich get richer''
phenomenon. Despite the generality of the proposed stochastic models, there are still
some unexplained phenomena, which may arise due to the limited size of networks such
as protein, e-mail, actor and collaboration networks. Such networks may in fact
exhibit an exponential cutoff in the power-law scaling, although this cutoff may only
be observable in the tail of the distribution for extremely large networks. We
propose a modification of the basic stochastic evolutionary model, so that after a
node is chosen preferentially, say according to the number of its inlinks, there is a
small probability that this node will become inactive. We show that as a result of
this modification, by viewing the stochastic process in terms of an urn transfer
model, we obtain a power-law distribution with an exponential cutoff. Unlike many
other models, the current model can capture instances where the exponent of the
distribution is less than or equal to two. As a proof of concept, we demonstrate the
consistency of our model empirically by analysing the Mathematical Research
collaboration network, the distribution of which is known to follow a power law with
an exponential cutoff.

\end{abstract}

\section{Introduction}

Power-law distributions taking the form
\begin{equation}\label{eq:power-law}
f(i) = C \ i^{- \tau},
\end{equation}
where $C$ and $\tau$ are positive constants, are abundant in nature \cite{SORN00}.
The constant $\tau$ is called the {\em exponent} of the distribution. Examples of
such distributions are: {\em Zipf's law}, which states that the relative frequency of
words in a text is inversely proportional to their rank, {\em Pareto's law}, which
states that the number of people whose personal income is above a certain level
follows a power-law distribution with an exponent between 1.5 and 2 (Pareto's law is
also known as the {\em 80:20 law}, stating that about 20\% of the population earn
80\% of the income) and {\em Gutenberg-Richter's law}, which states that  over a
period of time, the number of earthquakes of a certain magnitude is roughly inversely
proportional to the magnitude. Recently, several researchers have detected power-law
distributions in the topology of several networks such as the World-Wide-Web
\cite{BROD00,KUMA00b} and author citation graphs \cite{REDN98}.

\medskip

The motivation for the current research is two-fold. First, from a complex network
perspective, we would like to understand the stochastic mechanisms that govern the
growth of a network. This has lead to fruitful interdisciplinary research by a
mixture of Computer Scientists, Mathematicians, Statisticians, Physicists, and Social
Scientists \cite{ALBE01,DORO00c,KRAP00,LEVE01c,NEWM01,PENN02}, who are actively
involved in investigating various characteristics of complex networks such as the
degree distribution of the nodes, the diameter, and the relative sizes of various
components. These researchers have proposed several stochastic models for the
evolution of complex networks; all of these have the common theme of {\em
preferential attachment}--- which results in the ``rich get richer'' phenomenon
--- for example, where new links to existing nodes are added in
proportion to the number of links to these nodes currently present.

An extension of the preferential attachment model, proposed in \cite{DORO00c}, takes
into account the ageing of nodes so that a link is connected to an old node, not only
preferentially, but also depending on the age of the node: the older the node is the
less likely it is that other nodes will be connected to it. It was shown in
\cite{DORO00c} that if the ageing function is a power law then the degree
distribution has a phase transition from a power-law distribution, when the exponent
of the ageing function is less than one, to an exponential distribution, when the
exponent is greater than one. A different model of node ageing was proposed in
\cite{AMAR00} with two alternative ageing functions. With the first function the time
a node remains `active', i.e. may acquire new links, decays exponentially, and with
the second function a node remains active until it has acquired a maximum number of
links. Both functions were shown by simulation to lead to an exponential cutoff in
the degree distribution, and for strong enough constraints the distribution appeared
to be purely exponential. Another explanation of the cutoff, offered in
\cite{MOSS02}, is that when a link is created the author of the link has limited
information processing capabilities and thus only considers linking to a fraction of
the existing nodes, those that appear to be ``interesting''. It was shown by
simulation that when the fraction of ``interesting nodes'' is less than one there is
a change from a power-law distribution to one that exhibits an exponential cutoff,
leading eventually to an exponential distribution when the fraction is much less than
one.

\smallskip

Second, a motivation for this research is that the viability and efficiency of
network algorithmics are affected by the statistical distributions that govern the
network's structure. For example, the discovered power-law distributions in the web
have recently found applications in local search strategies in web graphs
\cite{ADAM01a}, compression of web graphs \cite{ADLE01} and an analysis of the
robustness of networks against error and attack \cite{ALBE00c,JEON01}.

\medskip

Despite the generality of the proposed stochastic models for the evolution of complex
networks, there are still some unexplained phenomena; these may arise due to the
limited size of networks such as protein, e-mail, actor and collaboration networks.
Such networks may in fact exhibit an exponential cutoff in the power-law scaling,
although this cutoff may only be observable in the tail of the distribution for
extremely large networks. The exponential cutoff is of the form
\begin{equation}\label{eq:exp-cutoff}
f(i) = C \ i^{- \tau} q^i,
\end{equation}
with $0 < q < 1$. The exponent $\tau$  in (\ref{eq:exp-cutoff}) will be smaller than
the exponent that would be obtained if we tried to fit  to the data a power law
without a cutoff, like (\ref{eq:power-law}). Unlike many other models leading to
power-law distributions, models with a cutoff can capture situations in which the
exponent of the distribution is less than or equal to two, which would otherwise have
infinite expectation.

\medskip

An exponential cutoff has been observed in protein networks \cite{JEON01}, in e-mail
networks \cite{EBEL02}, in actor networks \cite{AMAR00}, in collaboration networks
\cite{NEWM01,GROSS02b}, and is apparently also present in the distribution of inlinks
in the web graph \cite{MOSS02}, where a cutoff had not previously been observed. We
believe it is likely, in many such cases where power-law distributions have been
observed, that better models would be obtained with an exponential cutoff like
(\ref{eq:exp-cutoff}), with $q$ very close to one.

\medskip

The main aim of this paper is to provide a stochastic evolutionary model for a class
of networks like collaboration networks that result in asymptotic power-law
distributions with an exponential cutoff. This model also enables us to explain some
phenomena where the exponent is less than or equal to two. As with many of these
stochastic growth models, the ideas originated from Simon's visionary paper published
in 1955 \cite{SIMO55}. At the very beginning of his paper, in equation (1.1), Simon
observed that the class of distribution functions he was about to analyse can be
approximated by a distribution like (\ref{eq:exp-cutoff}); he called the term $q^i$
the {\em convergence factor} and suggested that $q$ is close to one. He then went on
to present his well-known model that yields power-law distributions like
(\ref{eq:power-law}), and which has provided the basis for the models rediscovered
over 40 years later. Simon gave no explanation for the appearance, in practice, of
the convergence factor.

\smallskip

In a previous paper \cite{FENN02}, we dealt with a related class of networks that exhibit
an exponential cutoff, such as protein interaction networks, in which after a protein is
chosen preferentially, say according to the number of other proteins it interacts with,
there is a small probability that this protein is discarded from the network. E-mail
networks and the web graph are further examples belonging to this class of network.
However, in this paper we consider other networks that behave differently, such as
collaboration and actor networks. Consider a collaboration network: after an author is
chosen preferentially, according to the number of collaborators he/she currently has,
there is a small probability that this author will become inactive, but he/she will not be
removed from the network. Inactive authors do not start new collaborations but their
existing collaborations still persist in the network. Possible reasons for inactivity may
be the finite time window of the data used or because an author retires from collaborative
writing.

\medskip

The rest of the paper is organised as follows. In Section~\ref{sec:urn} we present an
urn transfer model that extends Simon's model by allowing an author, chosen as
described above, to sometimes become inactive. We then derive the steady-state
distribution of the model, which, as stated earlier, follows an asymptotic power law
with an exponential cutoff like (\ref{eq:exp-cutoff}). In
Section~\ref{sec:collab-graph} we demonstrate that our model can provide an
explanation of the empirical distributions found in collaboration networks. Finally,
in Section~\ref{sec:concluding} we give our concluding remarks.

\section{An Urn Transfer Model for Collaboration Networks}
\label{sec:urn}

We now briefly present an {\em urn transfer model} for a stochastic process that emulates
the situation where balls (which might represent authors) become inactive with a small
probability, but still remain in the system. We assume that a ball in the $i$th urn has
$i$ pins attached to it (which might represent the author's collaborations). The model is
a variant of our previous model of exponential cutoff \cite{FENN02}, where balls are
discarded with a small probability.

\medskip

We assume a countable number of ({\em unstarred}) urns, $urn_1, urn_2, urn_3, \ldots
\ $ and a corresponding set of {\em starred} urns $urn^*_1, urn^*_2, urn^*_3, \ldots
\ $, where the latter contain the inactive balls. Initially all the urns are empty
except $urn_1$, which has one ball in it. Let $F_i(k)$ and $F^*_i(k)$ be the number
of balls in $urn_i$ and $urn^*_i$, respectively, at stage $k$ of the stochastic
process, so $F_1(1) = 1$, all other $F_i(1) = 0$ and all $F^*_i(1) = 0$. Then, at
stage $k+1$ of the stochastic process, where $k \ge 1$, one of two things may occur:

\renewcommand{\labelenumi}{(\roman{enumi})}
\begin{enumerate}
\item with probability $p$, $0 < p < 1$, a new ball (with one pin attached) is
inserted into $urn_1$, or

\item with probability $1 - p$ an urn is selected, with $urn_i$ being selected with
probability proportional to $i F_i(k)$, the number of pins it contains (or attached
to it), and a ball is chosen from the selected urn, $urn_i$ say; then,

\begin{enumerate}
\item with probability $q$, $0 < q \le 1$, the chosen ball is transferred to
$urn_{i+1}$, (this is equivalent to attaching an additional pin to the ball chosen
from $urn_i$), or

\item with probability $1 - q$ the ball chosen is transferred to $urn^*_i$ (this is
equivalent to making the ball inactive).
\end{enumerate}
\end{enumerate}
\smallskip

In terms of our model \cite{FENN02}, this means that instead of discarding a ball
from $urn_i$, say, the ball is transferred into the corresponding starred urn,
$urn^*_i$. A ball in a starred urn takes no further part in the stochastic process,
i.e. it does not acquire any further pins and so never moves from its urn. In
particular, balls in starred urns have no effect on the preferential choices made
during the continuation of the stochastic process.

\medskip

We could modify the initial conditions so that, for example, $urn_1$ initially
contained $\delta > 1$ balls instead of one. It can be shown that any change in the
initial conditions will have no effect on the asymptotic distribution of the balls in
the urns as $k$ tends to infinity, provided the process does not terminate with all
of the unstarred urns empty.

In order for this not to occur it is necessary  that, on average, more balls are
added to the system than become inactive. To ensure this we require $p > (1-p)
(1-q)$, see \cite{FENN02}. In practice this condition will nearly always hold, so
from now on we assume this. This condition implies that the probability
that the urn transfer process will {\em not} terminate with all the unstarred urns
being empty is positive.

\medskip

More specifically, the probability of non-termination is $1 -
\left( (1-p) (1-q) / p \right)^\delta$;
this is just the probability that the gambler's fortune will
increase forever \cite{ROSS83}.

\medskip

Following Simon \cite{SIMO55}, we now state the mean-field equations for the urn
transfer model. For $i > 1$ we have
\begin{equation}\label{eq:ss0}
E_k(F_i(k+1)) = F_i(k) + \beta_k \Big(q (i-1) F_{i-1}(k) - i F_i(k) \Big),
\end{equation}
where $E_k(F_i(k+1))$ is the expected value of $F_i(k+1)$ given
the state of the model at stage $k$, and
\begin{equation}\label{eq:betak}
\beta_k = \frac{1 - p}{\sum_{i=1}^k \ i F_i(k)}
\end{equation}
is the normalising factor.

\smallskip

Equation (\ref{eq:ss0}) gives the expected number of balls in $urn_i$ at stage $k+1$.
This is equal to the previous number of balls in $urn_i$ plus the probability of
adding a ball to $urn_i$ from $urn_{i-1}$ in step (ii)(a) minus the probability of
removing a ball from $urn_i$ in step (ii).

\smallskip

In the boundary case, $i = 1$, we have
\begin{equation}\label{eq:initial}
E_k(F_1(k+1)) = F_1(k) + p - \beta_k \ F_1(k),
\end{equation}
\smallskip
for the expected number of balls in $urn_1$, which is equal to the previous number of
balls in the first urn plus the probability of inserting a new ball into $urn_1$ in
step (i) of the stochastic process defined above minus the probability of removing a
ball from $urn_1$ in step (ii).

\smallskip

For starred urns, for $i \ge 1$, corresponding to (\ref{eq:ss0})
and (\ref{eq:initial}), we have
\begin{equation}\label{eq:star-urn1}
E_k(F^*_i(k+1)) = F^*_i(k) + (1-q) \beta_k i F_i(k).
\end{equation}

\medskip

In order to solve the equations for the model, we make the assumption that, for large
$k$, the random variable $\beta_k$ can be approximated by a constant (i.e. non-random)
value depending only on $k$. We do this by replacing the denominator in the definition of
$\beta_k$ by an asymptotic approximation of its expectation. We observe that
approximating $\beta_k$ in this way is essentially a {\em mean-field} approach
\cite{BARA99a}.

\medskip

Let  $\theta^{(k)}$ be the expected value of the average number of pins attached to a
ball in a starred urn at stage $k$. We have shown \cite{FENN02} that $\theta^{(k)}$
is bounded above by $1/(1 - q)$, so it is reasonable to make the assumption that
$\theta^{(k)}$ tends to a limiting value $\theta$ as $k$ tends to infinity. It is
easy to see that the total number of pins attached to the balls in the unstarred urns
(i.e. the active balls) at stage $k$ is asymptotically
\begin{displaymath}
(p + (1-p) q - (1-p) (1-q) \theta) k + O(1).
\end{displaymath}
Therefore, letting
\begin{equation}\label{eq:beta}
\beta = \frac{1-p}{p + (1-p) q - (1-p) (1-q) \theta},
\end{equation}
we see that $k \beta_k$ tends to $\beta$ as $k$ tends to infinity.

\medskip

If we now make the further assumption that
\begin{displaymath}
\theta^{(k)} = \theta + O(1 / k),
\end{displaymath}
then it is possible to show \cite{FENN02} that the expected value of $F_i(k)$ is
asymptotically proportional to $k$, i.e. $E(F_i (k)) / k$ tends to a limit $f_i$ as
$k$ tends to infinity. It similarly follows that $E(F^*_i (k)) / k$ tends to a limit
$f^*_i$.

\medskip

Following the derivation in \cite{FENN02}, we obtain
\begin{equation}\label{eq:ss2}
f_i = \beta \Big(q (i-1) f_{i-1} - i f_i \Big),
\end{equation}
for $i > 1$, and
\begin{equation}\label{eq:ss1}
f_1 = p - \beta f_1.
\end{equation}
The solution of these equations is
\begin{equation}\label{eq:ss5}
f_i = \frac{\varrho \ p \ \Gamma(1 + \varrho) \ \Gamma(i) \ q^i}{q \ \Gamma(i + 1 +
\varrho)} \ \sim \ \frac{C \ q^i}{i^{1 + \varrho}},
\end{equation}
where $\varrho = 1 / \beta$, $\Gamma$ is the gamma function \cite[6.1]{ABRA72} and
\begin{displaymath}
C = \frac{\varrho \ p \ \Gamma(1 + \varrho)}{q}.
\end{displaymath}

The asymptotic approximation to $f_i$ , i.e. (\ref{eq:ss5}), in the form
corresponding to (\ref{eq:exp-cutoff}) is obtained using Stirling's approximation
\cite[6.1.39]{ABRA72}.

\smallskip

For the starred urns, corresponding to (\ref{eq:ss2}) and (\ref{eq:ss1}), from
(\ref{eq:star-urn1}) we have, for $i \ge 1$,
\begin{equation}\label{eq:star-urn2}
f^*_i = (1-q) \beta i f_i.
\end{equation}
Thus the ratio of active balls in $urn_i$ to inactive balls in
$urn^*_i$ is
\begin{displaymath}
\frac{f_i}{f^*_i} = \frac{\varrho}{(1-q) i}.
\end{displaymath}

\smallskip

It follows that, for large $i$, the distribution of the balls is dominated by the
contents of the starred urns rather than the unstarred urns. Thus the distribution of
the total number of balls with $i$ pins is given by
\begin{equation}\label{eq:mixture}
f_i + f^*_i \sim \frac{C \ q^i}{i^\varrho} \left( \frac{1}{i} + \frac{1-q}{\varrho}
\right).
\end{equation}

\medskip

In the following section we will make use of the equation
\begin{equation}\label{eq:solve-pins2}
(1-p) (1 + \varrho) =  p \ F(1, 2; 2 + \varrho; q),
\end{equation}
where $F$ is the hypergeometric function \cite[15.1.1]{ABRA72}. This can be derived by
using (\ref{eq:ss5}) to obtain the sum of $i f_i$ for $i \ge 1$; this is just the
asymptotic value of the total number of pins attached to the balls in the unstarred urns
divided by $k$. However, from (\ref{eq:beta}) and the discussion preceding it, this sum is
also equal to $(1 - p)/ {\beta}$, i.e. $\varrho (1-p)$, see \cite{FENN02} for further
details.

\section{Collaboration Networks}
\label{sec:collab-graph}

As a proof of concept we will consider the Mathematical Research (MR) collaboration
network for which an exponential cutoff has been reported \cite{GROSS02b}. In our model it
is assumed that an author enters the network with a single collaboration, which could be
interpreted as a ``self-collaboration''. Thereafter, each time an author acquires a new
collaborator the corresponding ball is moved along to the next urn with an additional pin
attached to it. There is also a certain probability that an author becomes inactive.
Authors who are no longer active still remain part of the network, although they will not
be involved in any new collaborations.

\smallskip

We note that collaboration networks, together with some other types of network, like
protein and actor networks, are essentially undirected. So in our model a new
collaboration between two authors should be represented by two separate events, one
for each author. This would correspond to taking in pairs the events of attaching a
pin to a ball. We ignore this complication, but note that many of the models
proposed, for example for the web graph, similarly ignore the difference between
directed and undirected graphs (e.g. \cite{BARA99b}).


\medskip

We now examine in detail the degree distribution of the MR collaboration network. The
data for this was supplied to us by Jerry Grossman at the Department of Mathematics
and Statistics in Oakland University, Rochester \cite{GROSS02b}. In order to derive
the values for $\varrho$ and $q$, we performed a nonlinear regression on a log-log
transformation of the degree distribution of the MR collaboration network to fit the
equation
\begin{equation}\label{eq:regress}
y = a - \varrho \, x + \exp(x) \ln q + \ln(\exp(-x) + (1 - q)/ \varrho),
\end{equation}
corresponding to (\ref{eq:mixture}), where $a$ is a constant.

\smallskip

\begin{figure}[ht]
\centerline{\includegraphics[width=12cm,height=9.33cm]{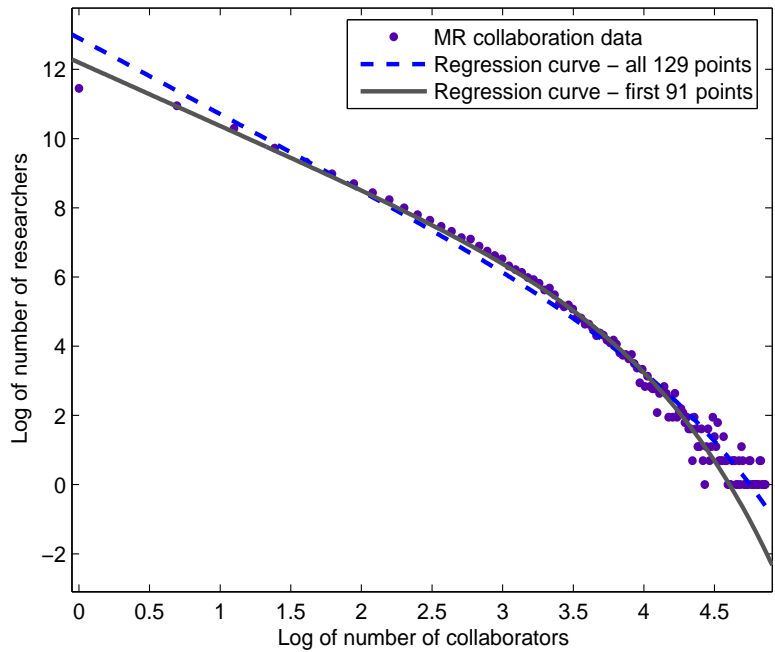}}
\caption{\label{fig:regress} Mathematical Research collaboration data}
\end{figure}

The results are shown in Figure~\ref{fig:regress}. The values of $\varrho$ and $q$
obtained from the regression of the complete MR data set (129 points) are $\varrho =
1.179$ and $q = 0.9658$.

\medskip

We next performed a stochastic simulation to test the validity of our model with respect
to the results of the regression on the original data set. In order to use this data for a
stochastic simulation of our model, we require the values for $p$ and $k$, using $\varrho$
and $q$ computed from the above regression.

\smallskip

We calculated a value for $k$ to use in simulating our stochastic model from:
\begin{equation}\label{eq:solve-balls}
\frac{balls_k + balls^*_k}{k} \approx p.
\end{equation}

This follows from the formulation of the model, where $balls_k$ and $balls^*_k$ stand for
the expected numbers of balls at stage $k$ in the unstarred and starred urns,
respectively.  The right-hand side of (\ref{eq:solve-balls}) is the limiting value of the
left-hand side as $k$ tends to infinity.

\smallskip

Similarly, from the formulation of the model, we have
\begin{equation}\label{eq:pins-star}
\frac{pins_k + pins^*_k}{k} \approx 1 - (1-p) (1-q).
\end{equation}
\smallskip

On using (\ref{eq:solve-balls}) and (\ref{eq:pins-star}), we obtain an alternative
equation for $p$, given by
\begin{equation}\label{eq:bf}
p \approx \frac{q}{bf + q - 1}
\end{equation}
where $bf = (pins_k + pins^*_k) / (balls_k + balls^*_k)$ is the branching factor.

\smallskip

From the data we see that the total number of researchers was 253339, and the total
number of collaborations was 992978. Using these values for $balls_k + balls^*_k$ and
$pins_k + pins^*_k$, respectively, gives us the branching factor for the original
data set as $bf = 3.9196$.

\smallskip

We can now obtain the values of $p$ and $k$. Computing $p$ from
(\ref{eq:solve-pins2}) gives $p = 0.3351$ and from (\ref{eq:bf}) gives $p = 0.2486$.
Using the first value of $p$ we obtain the alternative values for $k$ from
(\ref{eq:solve-balls}) or (\ref{eq:pins-star}) as  $k = 756010$ or $k = 1016083$,
respectively, and using the second value of $p$ gives us a value of $k = 1019170$
from either (\ref{eq:solve-balls}) or (\ref{eq:pins-star}).

\smallskip

We then carried out 10 simulation runs (a batch) of the stochastic process for the
three combinations of the values of $p, q$ and $k$. The results from the three
batches are shown in Table~\ref{table:collab1}. For each batch we report the average
output values for $balls_k + balls^*_k$, $pins_k + pins^*_k$ and $\varrho$. As a
further validation of our methodology, we computed the average number of balls in
each urn for each of the three batches, and performed a nonlinear regression, taking
into account all urns until an empty one was encountered. The values of $q$ and
$\varrho$ obtained from this regression are shown in the row following the results
for each batch.

\smallskip

For the first batch, it can be seen that the values of $balls_k + balls^*_k$ and
$\varrho$ are consistent with the data but the value of $pins_k + pins^*_k$ is less
consistent, since, in this case, we computed $k$ from (\ref{eq:solve-balls}). For the
second batch, it can be seen that the values of $pins_k + pins^*_k$ and $\varrho$ are
consistent with the data but the value of $balls_k + balls^*_k$ is less consistent,
since, in this case, we computed $k$ from (\ref{eq:pins-star}). Finally, for the
third batch, it can be seen that the values of $balls_k + balls^*_k$ and $pins_k +
pins^*_k$ are consistent with the data but the value of $\varrho$ is less consistent,
since $p$, computed from (\ref{eq:bf}), is less constrained than when it is computed
from (\ref{eq:solve-pins2}), which takes $\varrho$ into account. It is also evident
that value of $\varrho$ computed from the nonlinear regression on the urn values
from the simulation is, for all batches, below the value predicted from the
simulation.

\begin{table}[ht]
\begin{center}
\begin{tabular}{|l|r|r|r|r|r|r|} \hline
Simulation & $q$ & $p$ & $k$ & $balls_k + balls^*_k$ & $pins_k + pins^*_k$ & $\varrho$ \\
Data & $0.9658$ & $-$ & $-$ & $253339.0$ & $992978.0$ & $1.1790$ \\
\hline \hline
Batch 1 & $-$ & $0.3351$ & $756010$  & $253343.5$ & $738836.6$ & $1.1786$ \\
Regression & $0.9625$ & $-$ & $-$ & $-$ & $-$ & $1.0270$ \\ \hline
Batch 2 & $-$ & $0.3351$ & $1016083$ & $340592.2$ & $993041.2$ & $1.1795$ \\
Regression & $0.9641$ & $-$ & $-$ & $-$ & $-$ & $1.0530$ \\ \hline
Batch 3 & $-$ & $0.2486$ & $1019170$ & $254116.5$ & $993518.4$ & $0.9055$ \\
Regression & $0.9640$ & $-$ & $-$ & $-$ & $-$ & $0.8181$ \\
\hline
\end{tabular}
\end{center}
\caption{\label{table:collab1} Summary of simulations for parameters derived from the
full MR data set}
\end{table}
\medskip

We observe that there are problems in fitting power-law type distributions, due to
difficulties with non-monotonic fluctuations in the tail. (Another reason maybe the
sensitivity of the nonlinear regression to the cutoff parameter $q$.) In particular,
the presence of {\em gaps} in the distribution of balls in the urns is the main
manifestation of this problem. There is a {\em gap} in this distribution at $urn_i$
if there are no balls in $urn_i$ but there exists at least one ball in $urn_j$, where
$j > i$. We discussed this problem more fully in the context of a pure power-law
distribution in \cite{FENN03b}, and concluded that a preferable approach is to ignore
all data points from the first gap onwards. Evidence of the advantage of discarding
data points in the tail of the distribution was also given in \cite{GOLD04}, where
the more radical approach of using only the first five data points is suggested. In
the MR data set the first gap occurs at $i = 92$.

\smallskip

As a further test of the validity of the model, we created a truncated data set by
keeping only the first 91 data points of the MR data set. The regression curve, for
the first 91 points in the data set, is also shown in Figure~\ref{fig:regress}, where
the values for $\varrho$ and $q$ obtained from the regression are $\varrho = 0.8347$
and $q = 0.9438$.

\smallskip

Using these values for $\varrho$ and $q$ we obtained alternative values for $p$ and
$q$. Computing $p$ from (\ref{eq:solve-pins2}) gives $p = 0.2650$ and from
(\ref{eq:bf}) gives $p = 0.2443$. Using the value $p = 0.2650$ we obtain, from
(\ref{eq:solve-balls}) or (\ref{eq:pins-star}), the alternative values for $k$ as $k
= 955996$ or $k = 1035762$, respectively, and using the value $p = 0.2443$ gives us a
value of $k = 1037021$ from either (\ref{eq:solve-balls}) or (\ref{eq:pins-star}).

\smallskip

We then carried out 10 further simulation runs (a batch) of the stochastic process
for the three combinations of the values of $p, q$ and $k$, derived from the
truncated data set. The results from these further three batches are shown in
Table~\ref{table:collab2}.

\smallskip

For the first batch, it can be seen that the values of $balls_k + balls^*_k$ and
$\varrho$ are consistent with the data but the value of $pins_k + pins^*_k$ is less
consistent, although it is closer to its computed value compared to the value
$738836.6$ obtained from the previous simulations on the full MR data set. For the
second batch, it can be seen that the values of $pins_k + pins^*_k$ and $\varrho$ are
consistent with the data but the value of $balls_k + balls^*_k$ is less consistent,
although it is much closer to its computed value compared to the value $340592.2$
obtained from the previous simulations on the full data set. Finally, for the third
batch, it can be seen that the values of $balls_k + balls^*_k$ and $pins_k +
pins^*_k$ are consistent with the data but the value of $\varrho$ is less consistent,
although it is much closer to its computed value $0.8347$ compared to the value
$0.9055$ obtained from the previous simulations on the full data set; the latter is
further away from $1.179$. As for the full data set, it is also evident that value of
$\varrho$ computed from the nonlinear regression on the urn values from the
simulation is, for all batches, below the value predicted from the simulation.

\medskip

Overall the results show that the data is consistent with our model, and that the
results of the simulations better match the truncated data set. It is important to
note that small variations in $q$ obtained from the nonlinear regression may result
in relatively large variations in the regressed value of $\varrho$.

\begin{table}[ht]
\begin{center}
\begin{tabular}{|l|r|r|r|r|r|r|} \hline
Simulation & $q$ & $p$ & $k$ & $balls_k + balls^*_k$ & $pins_k + pins^*_k$ & $\varrho$ \\
Data & $0.9438$ & $-$ & $-$ & $253339.0$ & $992978.0$ & $0.8347$ \\
\hline \hline
Batch 1 & $-$ & $0.2650$ & $955996$  & $253585.3$ & $916594.8$ & $0.8353$ \\
Regression & $0.9402$ & $-$ & $-$ & $-$ & $-$ & $0.7273$ \\ \hline
Batch 2 & $-$ & $0.2650$ & $1035762$ & $274969.4$ & $993029.5$ & $0.8358$ \\
Regression & $0.9468$ & $-$ & $-$ & $-$ & $-$ & $0.8122$ \\ \hline
Batch 3 & $-$ & $0.2433$ & $1037021$ & $253840.0$ & $993041.8$ & $0.7681$ \\
Regression & $0.9428$ & $-$ & $-$ & $-$ & $-$ & $0.6975$ \\
\hline
\end{tabular}
\end{center}
\caption{\label{table:collab2} Summary of simulations for parameters derived from the
truncated MR data set}
\end{table}

\section{Concluding Remarks}
\label{sec:concluding}

We have presented an extension of Simon's classical stochastic process, which results
in a power-law distribution with an exponential cutoff. When viewing the stochastic
process in terms of an urn transfer model, the difference from the classical process
is that, after a ball is chosen on the basis of preferential attachment, with
probability $1-q$ the ball becomes inactive. By following the mean-field approach, we
derived the asymptotic formula (\ref{eq:mixture}), which shows that the distribution
of the number of balls in the urns approximately follows a power-law distribution
with an exponential cutoff.

\smallskip

Exponential cutoffs have been identified in protein, e-mail, actor and collaboration
networks, and possibly in the web graph \cite{MOSS02}; it is likely that exponential
cutoffs also occur in other complex networks. Here we have dealt with networks such
as collaboration and actor networks, where preferentially chosen authors/actors may
become inactive; in a previous paper (\cite{FENN02}) we have dealt with networks such
as protein and e-mail networks, where preferentially chosen proteins/e-mail accounts
may be discarded from the network. We demonstrated the applicability of our model
using data from the Mathematical Research collaboration network, thus showing that
our model offers a plausible explanation for certain processes that give rise to a
power-law distribution with an exponential cutoff.

\newcommand{\etalchar}[1]{$^{#1}$}

\end{document}